\documentclass{%
    optica-article%
}

\graphicspath{{Figures/}}
\usepackage[exponent-product=\cdot, separate-uncertainty=true]{siunitx}
\usepackage{booktabs}
\usepackage[version=4]{mhchem}
\usepackage{cleveref}

\usepackage{graphicx}
\usepackage{braket}

\usepackage{microtype}

\journal{opticajournal}
\articletype{Research Article}

\newcommand{\refcite}[2][\empty]{%
    \ifx#1\empty
        Ref.~\cite{#2}%
    \else%
        Ref.~\cite[#1]{#2}%
    \fi%
}
\newcommand{\mathpunc}[1][.]{
    \;\text{#1}
}

\newcommand{\diff}{\mathrm{d}\mspace{-1mu}}

\newcommand{\vrnum}[2]{No.~#1~-~#2}
\newcommand{\kawnum}[2]{No.~#1.#2}

\newcommand{\inputfigure}[3][width=\linewidth]{
    \begin{figure}
        \centering
        \includegraphics[#1]{#2}
        \caption{#3}\label{fig:#2}
    \end{figure}
}

\DeclareSIUnit{\atompercent}{at\%}
\DeclareSIUnit\bar{bar}
\DeclareSIQualifier{\RMS}{RMS}
\DeclareSIQualifier{\diameter}{\oslash}

\newcommand{\level}[3]{\ce{{}^{#1}#2_#3}}

\newcommand{\win}{\text{win}}
\newcommand{\ih}{\text{ih}}
\newcommand{\mode}{\text{mode}}
\newcommand{\FSR}{\text{FSR}}
\newcommand{\RT}{\text{RT}}
\newcommand{\cav}{\text{cav}}

\newcommand{\minimum}{\text{min}}
\newcommand{\nc}{\text{nc}}
\newcommand{\rem}{\text{rem}}
\newcommand{\cent}{\text{c}}

\newcommand{\EuYSO}{\ce{Eu{:}YSO}}

\begin{document}
\title{%
    Using slow light to enable laser frequency stabilization to a short, high-{Q} cavity
}
\author{%
    David Gustavsson,\authormark{1,*}
    Marcus Lindén,\authormark{1,2}
    Kevin Shortiss,\authormark{1}
    Stefan Kröll,\authormark{1}
    Andreas Walther,\authormark{1}
    Adam Kinos,\authormark{1}
    and
    Lars Rippe\authormark{1}
}
\address{
    \authormark{1}Lund University, Faculty of Engineering, Department of Physics, Box 188, 221~00 Lund, Sweden
    \authormark{2}Measurement Science and Technology, RISE Research Institutes of Sweden
}
\email{\authormark{*}david.e.gustavsson@gmail.com}
\date{\today}
\begin{abstract*}
    \bfseries
    State-of-the-art laser frequency stabilization is limited by miniscule length changes caused by thermal noise.
    In this work, a cavity-length-insensitive frequency stabilization scheme is implemented using strong dispersion in a \qty{21}{\mm} long cavity with a europium-ion-doped spacer of yttrium orthosilicate.
    A number of limiting factors for slow light laser stabilization are evaluated, including the inhomogeneous and homogeneous linewidth of the ions, the deterioration of spectral windows, and the linewidth of the cavity modes.
    Using strong dispersion, the cavity modes were narrowed by a factor \qty{1.6e5}, leading to a cavity linewidth of \(\qty{3.0}{\kHz}\) and a \(Q\) factor of \num{1.7e11}.
    Frequency stabilization was demonstrated using a cavity mode in a
    spectral transparency region
    near the center of the inhomogeneous profile, showing an overlapping Allan deviation below \num{6e-14} and a linear drift rate of \qty{3.66}{\Hz\per\s}.
    Considering improvements that could be implemented, this makes the europium-based slow light laser frequency reference a promising candidate for ultra-precise tabletop frequency stabilization.
\end{abstract*}

\section{Introduction}

Modern high precision measurement relies on accurate and stable clocks and frequency references.
Developments in these devices have allowed advances in experiments such as gravity wave detection~\cite{LIGO_2009, LIGO_2016, Li_2018} and tests of relativity~\cite{Eisele_2009, Chou_2010a, Delva_2017, Lu_2023}.
The best clocks to date are optical clocks~\cite{Jiang_2011,Nicholson_2015,Ludlow_2015}, where the frequency generating  element is a laser, locked long-term to atomic resonance frequencies.
For short-term frequency stability, the laser is locked to a reference cavity, such that the relative length stability of the cavity is translated into a relative frequency stability for the laser.

State-of-the-art reference cavities are typically on the order of tens of centimeters long and mechanically and thermally stabilized to the level where the limiting factor is Brownian motion in the atoms forming the mirrors~\cite{Numata_2004, Häfner_2015}.
A number of avenues are being pursued to mitigate this Brownian length uncertainty, including cryogenic cooling, crystalline mirror coatings, increased locking beam widths, and longer cavities.

The uncertainty in position of atoms in Brownian motion scales with the square root of the temperature~\cite{Einstein_1905}, meaning that reducing from room temperature (\qty{300}{\K}) to a liquid helium cryostat (on the order of \qty{3}{\K}) gives one order of magnitude reduction~\cite{Wiens_2014, Wang_2023}.
Further cooling is possible, but comes with new technical challenges.

Conventional stabilization cavities use amorphous multi-layer thin-film mirrors, deposited by ion beam sputtering.
While these have good optical properties, they can be improved on with respect to Brownian noise.
Specifically, epitaxially grown monocrystalline layers of semiconductor materials have shown between one and two orders of magnitude reduction in Brownian length uncertainty~\cite{Cole_2013, Kedar_2023}.

Since the Brownian noise is independent across the mirror, it can be averaged down by increasing the locking beam area, and scales with the reciprocal of the beam radius at the mirror~\cite{Numata_2004}.
Efforts into increasing the beam area involve operating cavities near instability, in near-planar or near-concentric configurations~\cite{Amairi_2013}.
This leads to a trade-off between beam area and sensitivity to mechanical noise and misalignment, which makes an order of magnitude increase from the \qty{1.5}{\mm} diameter beams studied in \refcite{Amairi_2013} improbable.

The relative frequency uncertainty depends on the relative length uncertainty, and so scales as the reciprocal of the cavity length.
Ultra-stable cavities up to \qty{50}{\cm} long have been demonstrated~\cite{Häfner_2015}, but an order of magnitude improvement in this respect, appears unlikely in the foreseeable future as it requires mechanically and thermally stabilizing a \qty{5}{\m} long cavity.

All of these techniques are approaching their current technical limits, and can each realistically provide about one or two orders of magnitude reduction from the current state-of-the-art.
In \refcite{Horvath_2022}, we presented another avenue to decrease the Brownian length uncertainty using the slow light effect.
By spectral tailoring, narrow frequency regions where the material is transparent (transmission windows) can be generated in the inhomogeneously broadened absorption profile of a cryogenically cooled rare-earth crystal, resulting in strong dispersion.
When the crystal is included in a cavity, this dispersion has the effect of increasing the cavity round-trip time, effectively making the cavity orders of magnitude longer without the associated increase in difficulty of mechanical stabilization.

This method is notably compatible with most of the alternative approaches:
It is by necessity performed at cryogenic temperatures, can utilize flat mirror cavities which allows large locking beam radii, and does not in principle preclude crystalline mirror coating.
For this reason slow light frequency stabilization offers several additional orders of magnitude reduction in length sensitivity.
It does, however, introduce new sources of frequency uncertainty from the degradation over time of the spectral structures in the spacer material.

In this work, we describe and evaluate a flat-mirror slow light cavity made from europium doped yttrium orthosilicate where the choice of material, the design of the cavity and the methods used are all optimized to ultimately enable ultra-stabilization while minimizing the previously identified sources of degradation.

\section{Background and theory}

As the group velocity \(v_g\) decreases due to increasing dispersion, variation in mode frequency \(\nu\) due to the variation in cavity length \(L\) scales as~\cite{Horvath_2022}
\begin{equation}
    \frac{\diff \nu}{\nu} = -\frac{\diff L}{L}\frac{v_g}{c_0} = -\frac{\diff L}{L}\frac{n_0}{n_0+\nu\frac{\diff n}{\diff \nu}}\mathpunc[,]
\end{equation}
with refractive index \(n\) varying approximately linearly with frequency around the host refractive index \(n_0\).
\(c_0\) is the speed of light in vacuum.
Evidently, strong dispersion \(\diff n/\diff\nu\), such as that generated in a narrow spectral window, greatly reduces the cavity's sensitivity to length variations.

To use this effect, mirror coatings are deposited on two opposite faces of a rare-earth-ion-doped crystal in which a spectral window is generated using optical pumping.
The resulting dispersion narrows the cavity modes by orders of magnitude, and the reflection of a frequency modulated locking beam is used to generate a Pound-Drever-Hall (PDH) error signal.
A feedback loop is established using this signal to lock the laser frequency to one of the cavity's resonance frequencies.

Because of the aforementioned reduction in sensitivity to length changes, the primary cause of drift in the resonance frequency of such a reference cavity will be spectral degradation of the transmission window through off-resonant excitation by the locking beam.
The ions are homogeneously broadened, and the ions near the edges of the window, whose homogeneous absorption profile reaches into the window, will be off-resonantly excited.
As these ions absorb the locking beam, they get optically pumped away and deform the window.
This part of the drift therefore depends on the locking beam intensity and the distance in frequency of the mode from the edges of the window, relative to the homogeneous linewidth of the ions.

In this work, we use a crystal of europium-doped yttrium orthosilicate (\EuYSO, \ce{Eu^3+{:}Y_2SiO_5}).
\EuYSO{} displays a good combination of parameters for slow light stabilization, with spectral structures remaining in excess of \qty{49}{days} at \qty{1.15}{\K}~\cite{Oswald_2018}, and homogeneous linewidths as narrow as hundreds of \unit{Hz}~\cite{Equall_1994}.
The long hyperfine lifetimes mean the spontaneous window drift, which has previously been observed in \ce{Pr^3+{:}Y_2SiO_5}~\cite{Horvath_2022}, is negligible, and the narrow homogeneous linewidth means the window can be made narrow while still displaying a low off-resonant excitation.

Furthermore, we reduce off-resonant excitation by using a flat-mirror cavity, which allows us to use a large locking beam area and therefore low intensity.
A flat mirror cavity has degenerate transversal modes, and the resonance frequency would typically be very sensitive to fluctuations in locking beam angle, but just as in the case of length changes this sensitivity is reduced by the strong dispersion, as derived in \cref{app:anglesensitivity}, and flat mirrors can therefore be used in the slow light stabilization scheme.
A large beam radius reduces the locking intensity and thereby the off-resonant excitation, and carries the additional benefit of averaging down the Brownian length uncertainty.

In order to further minimize drift caused by off-resonant excitation, we present a method to shift the cavity modes relative to the center frequency of the spectral transmission window by iteratively burning windows at different frequencies relative to the inhomogeneous profile.
This is done with the goal of placing a mode right at the window center frequency.
Locking to such a centered mode causes symmetric degradation at both edges of the window, eliminating first-order drift.
As an additional benefit, the total rate of degradation is minimized by minimizing the total off-resonant absorption.
As derived in \cref{app:pitvsmode}, moving the center frequency of the window \(\nu_\win\) will shift the mode frequency \(\nu_\mode\) relative to the window by a factor which is reduced about as much as the group velocity,
\begin{equation}
    \frac{\diff\nu_\text{mode}}{\diff\nu_\win} = -\frac{n_0}{n_g} +\frac{\pi\Gamma_\win}{2\Gamma_\ih}
\end{equation}
with group refractive index \(n_g\), host refractive index \(n_0\), window width \(\Gamma_\win\) and inhomogeneous linewidth \(\Gamma_\ih\).

To get a strong locking signal, the contrast between resonance and non-resonance needs to be maximized, which requires impedance matching.
This means the transmission of the front mirror should equal the effective losses of the cavity, including transmission of the back mirror and round-trip losses through the cavity.
If the front and back mirror reflectivity are \(R_1\) and \(R_2\) respectively, and the material has an absorption coefficient \(\alpha L\) and a round-trip length \(L_\RT\) such that the round-trip intracavity transmission is \(\exp(-\alpha L_\RT)R_1R_2\), the on-resonance reflectivity of the cavity in total is
\begin{equation}\label{eq:R_cav}
    R_\cav = R_1\left\lvert 1-\frac{1-R_1}{\sqrt{\frac{R_1}{R_2}\exp(\alpha L_\RT)}-R_1} \right\rvert^2 \mathpunc[.]
\end{equation}
Impedance matching means \(\alpha L_\RT = \ln(R_2/R_1) \Rightarrow R_\cav = 0\).
In a spectral transmission window, there will always be some off-resonant absorption from the ions near the edges of the window, and the narrower the window, the larger the absorption is in its center.
In this way, the window width, \(\Gamma_\win\), can be tuned to adjust the absorption in the center of the window, \(\alpha_\cent\), and seek impedance matching.
The center absorption \(\alpha_\cent\) for a square window of width \(\Gamma_\win\) in a material with homogeneous linewidth \(\Gamma_h\) in a region where the absorption outside the window is \(\alpha\) was shown in \refcite{Horvath_2022} to be
\begin{equation} \label{eq:alpha_cent}
    \alpha_\cent \approx \frac{2}{\pi}\frac{\Gamma_h}{\Gamma_\win}\alpha\text{.}
\end{equation}
As long as the total cavity losses are below impedance matching for some large \(\Gamma_\win\), such matching can be achieved by reducing \(\Gamma_\win\) until the desired \(\alpha_\cent\) is achieved.

\section{Experiments}

\subsection{Characterization}

\inputfigure{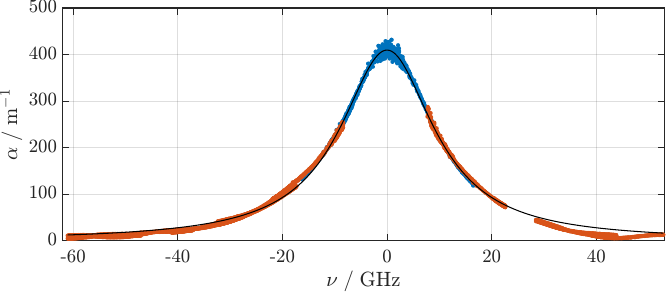}{
    Inhomogeneous profile of the \(\level{7}{F}{0}\to\level{5}{D}{0}\) transition in \qty{1}{\atompercent} \ce{Eu^3+{:}Y2SiO5}, with \(\nu_0 = \qty{516.85}{\THz}\) subtracted.
    Red and blue are measured with two different gain settings of the detector.
    The black line is a Lorentzian profile fit with full-width half-maximum linewidth \(\Gamma_\mathrm{i} = \qty{22}{\GHz}\) and peak height \(\alpha_0=\qty{410}{\per\m}\).
}

Good slow light frequency stabilization requires large dispersion, which means that a narrow spectral transmission window with sharp edges must be created within a region of large absorption.
Furthermore, the transmission window must be stable in time to enable frequency locking for an extended period.
The \level{7}{F}{0} transition in europium doped to \qty{1}{\atompercent} into YSO has a narrow homogeneous linewidth, and was measured in this crystal, using two-pulse photon echos, to be \qty{440}{\Hz} corresponding to a coherence time, \(T_2\), of \qty{725}{\micro\s}.
This narrow linewidth makes it possible to create windows with sharp edges.
Furthermore, the material has previously been shown to have extremely long hyperfine lifetimes, and no evident decay from hyperfine cross-relaxation was observed over our experiment durations.
The free spectral range of our cavity (length \(L=\qty{21}{\mm}\) and host refractive index \(n_0=1.8\)) is \(\nu_\mathrm{FSR}=c_0/\left(2n_0L\right)=\qty{4}{\GHz}\), and the inhomogeneous linewidth needs to be wide relative to this to ensure that there is a choice of window frequency within a region of large absorption.
We therefore used a high europium doping of \qty{1}{\percent}, which gave a full-width half-maximum inhomogeneous linewidth of \qty{22}{\GHz}, in line with the value reported in \refcite{Könz_2003}.
The inhomogeneous profile was measured in transmission using two detectors with different gain settings to capture the full dynamic range, and a Lorentzian profile was fitted as shown in \cref{fig:inhomogeneous}.
To compensate a calibration mismatch, evident from the region where the two data sets overlap, the low-gain data has been scaled down by \qty{8}{\percent}.

\begin{figure}
    \centering
    \hfill\includegraphics[width=0.65\textwidth]{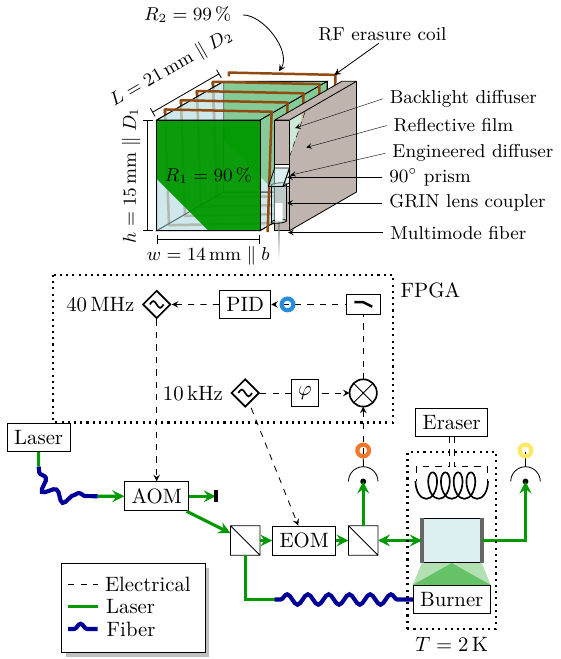}\hfill\includegraphics[width=0.33\textwidth]{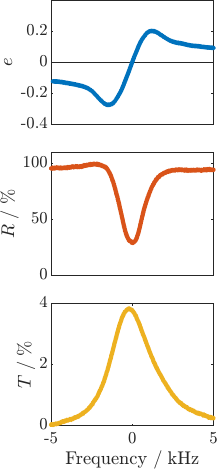}\hfill\\
    \caption{
        Top left: The crystal used in experiments, with dimensions and crystallographic axes marked.
        Rough geometry of the RF erasure coil, as well as the components of the side burner module.
        Bottom left: Schematic of the setup for stabilization experiments.
        Signal collection points are marked with colored circles, matching the respective trace on the right.
        The relative geometry of the components inside the cryostat is more accurate in the top left graphic.
        Right: PDH error signal (\(e\)), cavity reflection (\(R_\cav\)) and transmission (\(T_\cav\)), averaged over several frequency-chirped readouts.
        The transmission linewidth is \(\Delta\nu=\qty{3}{\kHz}\).
    }\label{fig:setup}
\end{figure}

\inputfigure{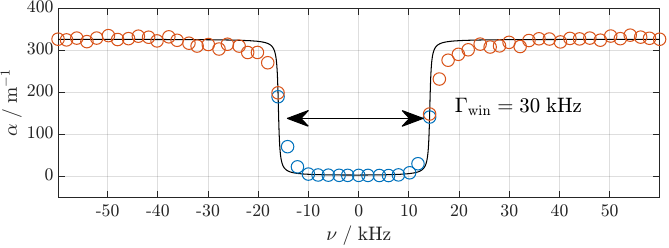}{ 
    A narrow spectral transmission window of width \(\Gamma_\win=\qty{30}{\kHz}\).
    Absorption measured by transmission of long, spectrally narrow pulses using two detectors with different photodetector gain settings (blue: low gain, red: high gain).
    In black is the theoretical absorption profile of a square-population window, a sum of arctangent curves, derived in \refcite{Horvath_2022}.
    The discrepancy between the data and the theoretical model is explained by a slight over-burning of ions in the edges of the window.
}

\inputfigure{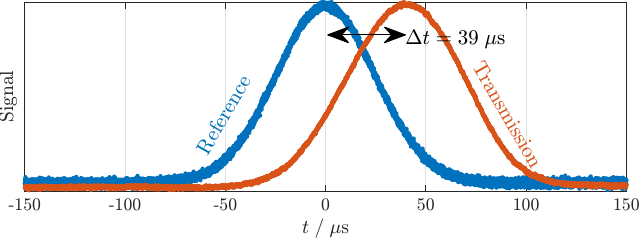}{ 
    A Gaussian pulse propagated through the crystal in the middle of a \qty{30}{\kHz} wide spectral window near the center of the inhomogeneous profile (blue: reference, red: transmitted signal), showing the time delay for light propagating in the transmission window, \(\Delta t = \qty{39}{\us}\).
    The signals have been separately normalized, the vertical scale is arbitrary.
}

Through spectral tailoring, narrow transmission windows were burned into the absorption profile using optical pumping.
Repeated sechscan pulses (a linearly chirped square pulse bookended by two halves of a secant hyperbolicus pulse, introduced as HSH in \refcite{Tian_2011}) were used, in order to achieve sharp transmission window edges.
To avoid standing wave patterns and perform spectral tailoring evenly throughout the length of the crystal, a diffuse and wide burning beam was injected orthogonally to the probing axis.
A fiber-coupled sideburner was designed for this purpose, illustrated in \cref{fig:setup} (top left), and described further in \cref{app:experimentaldetails}.
By optimizing the optical pumping sequence, we were able to create windows as narrow as \qty{30}{\kHz}, see \cref{fig:pit}.
The absorption profile as a function of frequency was obtained by sending in long square pulses at different frequencies, which only probed a narrow frequency range.
To cover the large dynamic range of the signal, two detectors with different gain settings were used.
The dispersion was measured by sending a Gaussian pulse through the center of the window (\cref{fig:slow_pulse}), resulting in a pulse delay of \qty{39}{\us} corresponding to a group velocity of \qty{540}{\m\per\s} and a group refractive index of \num{5.6e5}.
These measurements were performed through an uncoated part of the crystal.

A cavity was constructed by depositing reflective coatings on the front and back face of the crystal, with \qty{90}{\percent} and \qty{99}{\percent} reflectivity respectively.
For locking experiments, the setup illustrated in \cref{fig:setup} was used.
PDH locking was established using a digital laser servo described in \refcite{Pomponio_2020}.
The shape of the measured reflection and transmission signals, as well as the PDH locking signal, when probed with a linearly chirped laser is shown in \cref{fig:setup} (right).
The full-width half-maximum mode width was measured in transmission to \(\Delta\nu=\qty{3}{\kHz}\), \num{1.61e+05} times narrower than the non-dispersive cavity linewidth.
The experimental quality factor for this cavity is \(Q=\nu_0/\Delta\nu=\num{1.7e+11}\), which is extremely high for a cavity like this one.
For comparison, the \(Q\) factor without dispersion is \num{1.06e+06}.
As derived in \refcite{Horvath_2022}, the minimum linewidth for an impedance matched slow light cavity is \(\Delta\nu_\minimum = 2\Gamma_h = \qty{880}{\Hz}\), which leads to a maximum theoretical \(Q\) value of \num{5.9e+11} for this particular cavity.

\subsection{Stabilization}

\begin{figure}
    \centering
    \includegraphics[width=\textwidth]{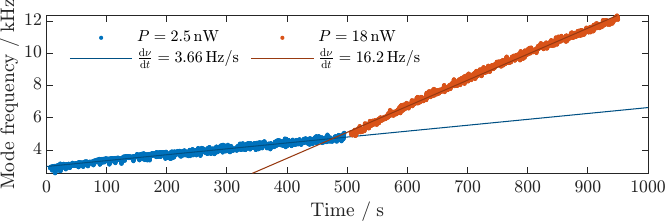}\\ 
    \includegraphics[width=\textwidth]{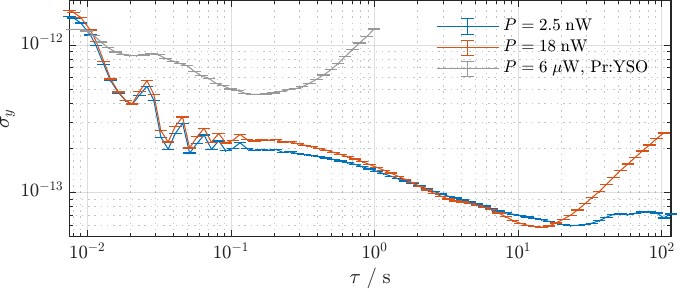}\\ 
    \caption{%
        Top:
        Long term drift of two \qty{500}{\s} locks in a \qty{40}{\kHz} wide window, where the locking power is increased after the first \qty{500}{\s}.
        The mode frequency is retrieved from the servo feedback signal.
        Bottom:
        Overlapping Allan deviation with linear drift subtracted, for two locks with different power in the europium doped crystal, compared to a praseodymium doped crystal (calculated from data reported in \refcite{Horvath_2022}).
        The measured noise floor is \(\sigma_y=\num{6e-14}\).
    }\label{fig:lock}
\end{figure}

Pound-Drever-Hall stabilization was established to a mode inside a \qty{40}{\kHz} wide window over \qty{500}{\s} intervals.
In order to measure the power induced drift from off-resonant excitation, a low power (approximately \qty{2.5}{\nW}) lock was followed by one at \num{7} times higher power.
One such lock is shown in \cref{fig:lock} (Top), where the drift rate at low power was measured to be \qty{3.66}{\Hz\per\second} when locking to a mode \qty{2}{\kHz} above the center frequency.
As discussed in \refcite{Horvath_2022}, this drift is likely from asymmetric off-resonant excitation of ions near the window edges, which causes the mode to move relative to the center of the window.
\Cref{fig:lock} (Bottom) shows the overlapping Allan deviation of these locks, with linear drift subtracted, as suggested in \refcite{Riley_2008}, compared to the lock to a praseodymium doped YSO crystal reported in \refcite{Horvath_2022}.
The laser noise floor is \(\sigma_y = \num{6e-14}\), at averaging times above \qty{10}{\s}, below which it is likely limited by fiber noise.
This is an order of magnitude improvement over the praseodymium value, using \num{300} times less locking power.

In order to lock to a mode near the center frequency of the transmission window, windows were created at several frequencies across the inhomogeneous absorption profile and the corresponding mode frequency measured (\cref{fig:lockvspit}).
Between each measurement, the structures were reset using RF erasure, evaluated separately in \refcite{Linden_2024b}.
A least squares fit was performed of \cref{eq:numode} (\cref{app:pitvsmode}), with the mode offset \(r\) and the peak absorption \(\alpha_0\) as fitting parameters, to the measured mode frequencies.
The best fit lines are marked in black in \cref{fig:lockvspit}.
The fit predicted a peak absorption of \qty{353}{\per\m}, somewhat lower than the \qty{410}{\per\m} measured in transmission (\cref{fig:inhomogeneous}).
This discrepancy could potentially be explained by imperfect spectral tailoring.
The overall behavior is approximately as predicted, so iteratively moving the window in steps of \(\Delta\nu_\win\) allows one to get a mode within \(\left(n_0/n_g\right)\Delta\nu_\win\) of the center of the window.
It is not unrealistic, in future work, to move the window frequency in steps of \(\Delta\nu_\win<\qty{1}{\kHz}\), giving \(\Delta\nu_\text{mode}<\qty{10}{\mHz}\).
As we have previously shown that the power-induced drift rate scales roughly linearly with mode distance from the center of the window~\cite{Horvath_2022}, this improvement should reduce the power-induced drift rate -- the dominant drift in this work -- by over five orders of magnitude, compared to \cref{fig:lock}.

\inputfigure{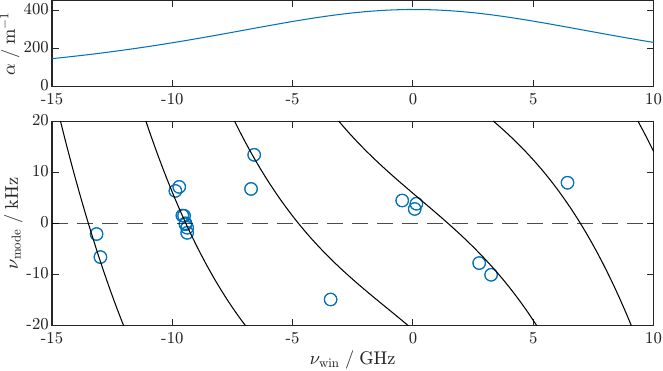}{ 
    Bottom:
    Mode frequency relative to window center frequency \(\nu_\mode\) vs window frequency relative to absorption peak frequency \(\nu_\win\) in \qty{40}{\kHz} wide spectral windows.
    Best fit of the theoretical model derived in \cref{app:pitvsmode} is marked with lines.
    The horizontal dashed line corresponds to the center of the window, where the drift is theoretically zero to first order.
    Top: Lorentzian fit of the inhomogeneous profile in \cref{fig:inhomogeneous}, as a visual guide.
}

\inputfigure{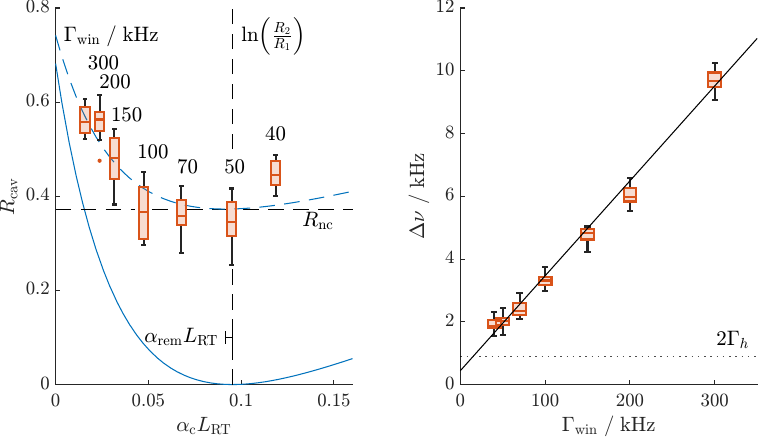}{
    Left:
    Reflectivity at resonance for sequentially widening windows, labeled with the respective window width.
    The horizontal position of each box is the theoretically predicted off-resonant absorption for a perfectly burned window of that width.
    Any residual absorption from remaining population or scattering losses will correspond to a horizontal shift.
    In dashed blue is a least-squares fit of the data with two fit parameters:
    a non-coupling portion \(R_\mathrm{nc}=\qty{37}{\percent}\) shifting the fit upwards and a remaining round-trip loss of \(\alpha_\rem L_\RT = \num{0.0038}\) shifting it vertically.
    Right:
    Cavity mode linewidth vs window width for the same modes, and a straight-line fit with a slope of \num{0.03}.
    The dotted line represents the theoretical minimum linewidth \(2\Gamma_h\), derived in \refcite{Horvath_2022}.
}

To verify impedance matching in the cavity, a \qty{40}{\kHz} wide window was burned near the center of the inhomogeneous profile, and then widened in steps without intermediate erasure, measuring the reflection and transmission of the cavity mode.
\Cref{eq:alpha_cent} relates the window width to an expected centre-of-window absorption \(\alpha_\cent\), which decreases with increasing window width.
\Cref{fig:impedance_matching} shows the measured on-resonance cavity reflectivity against \(\alpha_\cent\).
\Cref{eq:R_cav} was adapted into a fitting model with fitting parameters \(R_\nc\) and \(\alpha_\rem\),
\begin{equation} \label{eq:R_cavfit}
    R_\cav = R_\nc + (1-R_\nc)R_1\left\lvert 1- \frac{1-R_1}{\sqrt{\frac{R_1}{R_2}\exp\left(\left(\alpha_\cent+\alpha_\rem\right)L_\RT\right)}-R_1}\right\rvert^2
\end{equation}
where \(R_\nc\) represents a non-coupling portion of the intensity that is immediately reflected due to transversal mode mismatch, and \(\alpha_\rem\) represents remaining intracavity losses after compensating for off-resonant absorption \(\alpha_\cent\).
This model was fit to the data using least-squares fitting, giving \(R_\nc = \qty{37}{\percent}\) and \(\alpha_\rem L_\RT = \num{0.0038}\).
If \(\alpha_\rem L_\RT\) is entirely due to residual, unburned population, this population is \(\alpha_\rem/\alpha_0 = \num{0.03}\).
The fact that the cavity reflectivity increases with window width for the widest windows confirms that intra-cavity losses are low enough for impedance matching to occur in windows around \qty{50}{\kHz} wide.
The mode width, shown in \cref{fig:impedance_matching} (Right), increases roughly linearly with window width, as derived in \refcite{Horvath_2022}:
\begin{equation}
    \Delta\nu \approx \Gamma_h+\frac{\pi\left(2-R_1-R_2\right)}{4\alpha_0 L}\Gamma_\win \mathpunc[.]
\end{equation}
The measured slope is \num{0.03}, indicating that the finesse is \num{3} times lower than predicted in theory.
The linear relationship between window width and mode width is nevertheless satisfactory.

\section{Conclusion}

We have designed and tested a highly doped \ce{Eu{:}YSO} cavity and demonstrated that it is a promising candidate for ultra-stable slow light frequency stabilization.
The homogeneous linewidth is narrow enough that sharp spectral windows down to \qty{30}{\kHz} wide could be burned without excessive off-resonant absorption, and the resulting slow light effect on the order of \num{5e5} enabled extremely narrow cavity modes, with a \(Q\) value exceeding \num{e11}.
We showed that the spectral structures were stable down to the \unit{\Hz\per\s} level while locking to a mode \qty{2}{\kHz} from the center of the window.
Frequency stabilization was established for minutes at a time, with a measured noise floor of \(\sigma_y=\num{6e-14}\).
Determining the true achievable frequency stability of this reference will require more careful control of experimental conditions.

We have presented a technique by which the mode frequency can be tuned to the -- to first order -- drift-free position in the middle of the window, where power induced drift is no longer limiting.
This is made possible by the wide inhomogeneous linewidth of the material.
We have also demonstrated that, by choosing the window width appropriately, the losses in the cavity can be satisfactorily controlled to achieve impedance matching.

The slow light effect in \EuYSO{} potentially represents over five orders of magnitude reduction in sensitivity to length fluctuations, and does not conflict with other measures such as cryogenic cooling, crystalline mirrors, and increased locking beam radii.
Combining these measures has the potential to greatly reduce Brownian noise, the current limiting factor to optical frequency stability.

\section{Acknowledgements}

The authors would like to thank Marco Pomponio at NIST, who developed the digital servo and signal generator used, and provided much useful information and timely software updates.

This research was supported by
the Swedish Research Council (Grants \vrnum{2016}{05121}, \vrnum{2019}{04949}, and \vrnum{2021}{03755}),
the Knut and Alice Wallenberg Foundation (Grant \kawnum{2016}{0081}),
the Wallenberg Center for Quantum Technology funded by the Knut and Alice Wallenberg Foundation (Grant \kawnum{2017}{0449}),
the European Union EMPIR program (NEXTLASERS),
the  European Union FETFLAG program (SQUARE),
and the Fund of the Walter Gyllenberg Foundation.

\bibliography{references}

\newpage
\appendix
{\noindent\huge Appendix}

\section{Experimental details}\label{app:experimentaldetails}

The slow light stabilization scheme was realized in an yttrium orthosilicate (YSO, \ce{Y2SiO5}) crystal manufactured by Scientific Materials.
The crystal was doped to \qty{1}{\atompercent} with europium ions in natural abundance -- approximately equal parts \ce{{}^{151}Eu^3+} and \ce{{}^{153}Eu^3+}.
The sample's dimensions are \(\qty{14}{\mm}\times\qty{15}{\mm}\times\qty{21}{\mm}\) (optical axes \(D_1 \times b \times D_2\)).
The crystal was mechanically polished to minimize the wedge angle between the mirror faces (\qty{8}{\micro\radian}), and then superpolished by Ion Beam Forming (\qty{1}{\nm\RMS} over a \qty{1}{\mm} diameter beam, \qty{1.37}{\micro\radian} wedge angle).
In the polishing process, a spatial inhomogeneity in the refractive index was discovered which meant we could either make both mirror faces separately flat or the high-reflectivity face absolutely flat and the cavity optically flat in transmission.
We selected the second option, and the reported flatness numbers are measured through the crystal.
Finally, the front and back faces were coated to \qty{90}{\percent} and \qty{99}{\percent} reflectivity respectively using Ion Beam Sputtering.
A corner was left uncoated for transmission-mode experiments.

For the locking experiments, the crystal was placed on a cradle, vibrationally isolated from the rest of the setup by a \qty{3}{\mm} thick sheet of open-cell polyimide foam.
Adjacent to the crystal, but separated from it by a vacuum gap, were a tunable resonant RF antenna for hyperfine erasure, described in \refcite{Linden_2024a}, and a sideburner assembly.
The sideburner was designed to effectively perform optical pumping throughout the crystal from the side, avoiding standing wave effects.
The design is shown in \cref{fig:setup} (top left).
The burning light was fed into the cryostat via a multimode optical fiber.
On the output, the light was collimated by a GRIN lens, reflected by a \qty{90}{\degree} prism, and onto an engineered diffuser which spread the light out into a vertical line.
The light was then scattered through a diffusive piece of a commercially available LED backlight module, which was clad in a reflective film on all sides but the one facing the crystal.
This way, the crystal was evenly illuminated by a wide, diffuse beam.

All experiments were performed using a Coherent 699-21 ring laser with rhodamine 6G dye.
This laser was stabilized to an external ultra-low expansion glass (ULE) cavity and the linewidth is estimated to be on the order of a few tens of \unit{\Hz}.
The ULE cavituy was temperature stabilized close to its zero temperature coefficient point, which gives a low drift rate~\cite{Alnis_2008}.
The light was passed through a \qty{20}{\m} optical fiber, which on short time scales broadens it to approximately \qty{1}{\kHz}~\cite{MA_1994}, before entering the experimental table.
At the crystal, the locking beam had a diameter of approximately \qty{1}{\mm}.

The crystal was temperature stabilized to \qty{1.6}{\K} in a closed cycle cryostat from MyCryoFirm with a \qty{4}{\K} pulse-tube cooled stage and a secondary Joule-Thomson cooled stage with a gas cell for the sample.
The crystal was thermally connected to the cell through an injection of \qty{0.1}{\milli\bar} helium gas.

In order to restore the spectral profile between experiments, the hyperfine ground states were scrambled using an RF erasure technique described and evaluated in \refcite{Linden_2024b}.
An RF magnetic field was generated in a coil surrounding the crystal, at the resonance of a tunable resonance circuit, which was tuned to the \(\ket{\pm1/2}\to\ket{\pm3/2}\) and \(\ket{\pm3/2}\to\ket{\pm5/2}\) transitions in the ground state of each of the two isotopes -- four transitions in total -- performing repeated \(\pi/2\) sweeps for each one in sequence, although for \(\ket{\pm3/2}\to\ket{\pm5/2}\) in \ce{^153Eu^3+} a \(\pi/2\) pulse area could not be achieved due to low oscillator strength and limitations in pulse length and power.

Frequency stabilization was achieved with PDH locking using a digital laser servo and signal generator implemented on a Field Programmable Gate Array (Koheron,~alpha-250), marked FPGA in \cref{fig:setup} (bottom left).
The software is described in \refcite{Pomponio_2020}.
The FPGA generated a \qty{40}{\MHz} signal to drive a frequency shifting acousto-optic modulator (Isomet,~M1201-SF40).
It also acted as a local oscillator supplying a \qty{10}{\kHz} signal to a fiber-coupled electro-optic modulator (Jenoptik,~PM594), which generated side bands at a modulation of depth \(\beta\approx1.08\) in line with the optimum derived in \refcite{Black_2001}.
The reflected light off the crystal was detected by a photo diode (Hamamatsu,~S5973-02) connected to a transimpedance amplifier (NF,~SA-606F2) and fed back into the FPGA, where a digital phase detector and low pass filter converted the modulation into an error signal.
This error signal was then fed to a digital PID servo controlling a Numerically Controlled Oscillator (NCO) connected to a DAC.
This signal shifted the frequency out of the acousto-optic modulator accordingly.

The feedback signal was used to determine linear drift, as well as overlapping Allan deviation using the Matlab function \verb|allan_overlap|~\cite{Matlab_allan_overlap}.
To uncouple the power-induced drift from signal power effects, a neutral density filter was inserted before the reflection detector for the higher intensity experiments which compensated the increased power, meaning that only the drift rate, and not the noise, should be affected.

\section{The effect of window frequency on relative mode frequency}\label{app:pitvsmode}

In this appendix we derive the mode frequency relative to the center of a spectral transmission window, as a function of that window's center frequency relative to the inhomogeneous absorption profile.
The resonance condition for transmission through a cavity of length \(L\) is
\begin{equation}
    2L = q\lambda
\end{equation}
with resonant wavelengths \(\lambda\) and an integer \(q\).

\(\lambda = \frac{c_0}{n\nu}\) gives
\begin{equation}\label{eq:resonance}
    q = \frac{2nL\nu}{c_0}
\end{equation}

We define the frequency for the peak of the inhomogeneous profile \(\nu_0\), a window center frequency \(\nu_\win\) as the offset from \(\nu_0\) to the center of the window, and a mode frequency \(\nu_\mode\) which is the frequency of the mode relative to the center of the window.
The absolute mode frequency is then \(\nu = \nu_0+\nu_\win+\nu_\mode\).
Similarly, we write the refractive index \(n\) as the sum of the host refractive index \(n_0\), a contribution \(\Delta n_\ih(\nu_\mathrm{win})\) from the inhomogeneous profile and a contribution \(\Delta n_\win(\nu_\mode) = n'\nu_\mode\) from the linear dispersion with coefficient \(n'\) in the window.
This allows us to rewrite \cref{eq:resonance} as
\begin{equation}
    \left(1+\frac{\Delta n_\ih}{n_0}+\frac{\Delta n_\win}{n_0}\right)\left(1+\frac{\nu_\win}{\nu_0}+\frac{\nu_\mode}{\nu_0}\right) =
    \frac{q}{\nu_0}\frac{c_0}{2n_0L} =
    \frac{q\nu_\FSR}{\nu_0} \mathpunc[.]
\end{equation}

Here \(\nu_\FSR\) is the free spectral range of the cavity without the presence of the atomic line.
Recognizing that each of the non-unity terms in the left-hand side is much smaller than \(1\), their second-degree combinations can be neglected, which leaves
\begin{equation}\label{eq:negligent}
    1+\frac{\Delta n_\ih}{n_0}+\frac{\Delta n_\win}{n_0}+\frac{\nu_\win}{\nu_0}+\frac{\nu_\mode}{\nu_0} = \frac{q\nu_\FSR}{\nu_0}\mathpunc[.]
\end{equation}

We introduce an integer \(\Delta q\) and a real number \(r\in\left[-0.5,\,0.5\right)\) such that
\begin{equation}
    q\nu_\FSR = \nu_0+\left(\Delta q+r\right)\nu_\FSR\mathpunc[.]
\end{equation}
This amounts to counting the number of free spectral ranges from the cavity mode nearest to \(\nu_0\), instead of the absolute mode number.
\Cref{eq:negligent} simplifies to
\begin{equation}
    \frac{\Delta n_\ih}{n_0}+\frac{\Delta n_\win}{n_0}+\frac{\nu_\win}{\nu_0}+\frac{\nu_\mode}{\nu_0} = \left(\Delta q + r\right)\frac{\nu_\FSR}{\nu_0}\mathpunc[.]
\end{equation}
\begin{equation}
    \Delta n_\ih\nu_0 + n_0\nu_\win + \nu_\mode\left(n_0+n'\nu_0\right) = (\Delta q + r)n_0\nu_\FSR
\end{equation}

We separate out \(\nu_\mode\),
\begin{equation}\label{eq:brokenout}
    \nu_\mode = \frac{(\Delta q + r)n_0\nu_\FSR-n_0\nu_\win-n_\ih(\nu_\win)\nu_0}{n_0+n'\nu_0}
\end{equation}

From \refcite{Horvath_2022}, we have the group refractive index
\begin{equation}\label{eq:n_g}
    n_g = n_0 + n'\nu_0\approx \frac{c_0\alpha}{\pi^2\Gamma_\win} \mathpunc[.]
\end{equation}

Using the complex susceptibility of a Lorentzian absorption line (\refcite[Chapter 2.4]{Siegman_1986}), the inhomogeneous profile's contributions are derived to \begin{equation}\label{eq:susc_inhomo}\begin{gathered}
    \alpha = \frac{\alpha_0\Gamma_\ih^2}{4\left(\nu_\win^2+\frac{\Gamma_\ih^2}{4}\right)} \\
    n_\ih(\nu_\win)=-\frac{\nu_\win\Gamma_\ih\alpha_0c_0}{8\pi\nu_0\left(\nu_\win^2+\frac{\Gamma_\ih^2}{4}\right)}
\end{gathered}\end{equation}

Combining \cref{eq:brokenout,eq:n_g,eq:susc_inhomo} gives
\begin{equation}\label{eq:numode}
    \nu_\mode =
    (\Delta q + r)\frac{n_0}{n_g}\nu_\FSR-
    \frac{n_0}{n_g}\nu_\win+
    \frac{\pi\Gamma_\win}{2\Gamma_\ih}\nu_\win
\end{equation}

The three terms in \cref{eq:numode} predict three effects on the mode frequency as the window is moved.
The first is an effective free spectral range scaled down by the slow light factor \(n_0/n_g\) (note that \((\Delta q + r)\) takes values with integer spacing).
The second effect is that as \(\nu_\win\) changes, the change in \(\nu_\mode\) is that change scaled down by the same slow light factor.
The final term predicts a distortion due to the presence of the inhomogeneous line itself.
This effect does not depend on the peak absorption \(\alpha_0\), but the slope addition scales with the ratio between the window width and the inhomogeneous linewidth.
In our experiment, this makes the last term about \qty{10}{\percent} as large as the second.

\section{Angle sensitivity}\label{app:anglesensitivity}

Flat mirror cavities have degenerate transversal modes, so any fluctuations in angle of incidence are translated into shifts in mode frequency.
In this appendix we derive an expression for this dependence which includes dispersion in the cavity, showing that strong dispersion heavily reduces the sensitivity.

The phase accumulated for a light wave of frequency \(\nu\) traveling a distance \(z\) through a crystal with refractive index \(n\) is
\begin{equation}
    \Delta\phi(z) = \frac{2\pi}{c_0}n\nu z \mathpunc[.]
\end{equation}
From Snell's law and simple trigonometry, the single-pass path length through a crystal of length \(L\) with refractive index \(n\) for a beam with incidence angle \(\theta\) from vacuum is
\begin{equation}
    z = L\sqrt{1-\frac{1}{n^2}\sin^2\theta} \mathpunc[,]
\end{equation}
so the single-pass phase is
\begin{equation}
    \Delta\phi = \frac{2\pi}{c_0}n\nu L\sqrt{1-\frac{1}{n^2}\sin^2\theta} \mathpunc[.]
\end{equation}

Staying resonant as the angle varies necessarily implies a constant \(\Delta\phi\).
This means we can equate the expression with one for normal incidence and unshifted frequency, where \(n=n_0\), \(\nu=\nu_0\) and \(\theta=0\):
\begin{equation}
    n\nu\sqrt{1-\frac{1}{n^2}\sin^2\theta} = n_0\nu_0 \mathpunc[.]
\end{equation}
We introduce \(\nu_\mode = \nu - \nu_0\) and assume linear dispersion, \(n = n_0 + n'\nu_\mode\).
Furthermore we assume a small incident angle to simplify \(\sqrt{1-x}\approx1-x/2\)
\begin{equation}
    \left(1+\frac{n'\nu_\mode}{n_0}\right)\left(1+\frac{\nu_\mode}{\nu_0}\right)\left(1-\frac{\sin^2\theta}{2(n_0+n'\nu_\mode)^2}\right) = 1
\end{equation}

The non-unity terms in the parentheses are much smaller than \(1\), so higher order terms can be neglected, leaving
\begin{equation}
    \left(n_0 + n'\nu_0\right)\frac{\nu_\mode}{\nu_0} = n_0\frac{\sin^2\theta}{2n_0^2}
\end{equation}
and finally
\begin{equation}
    \frac{\nu_\mode}{\nu_0} = \frac{n_0}{n_0+\nu_0\frac{\diff n}{\diff \nu}}\frac{\sin^2\theta}{2n_0^2} \mathpunc[.]
\end{equation}
Just like with length changes, the angle sensitivity scales with the group refractive index \(n_g = n_0+\nu\frac{\diff n}{\diff\nu}\).

\end{document}